\documentclass[11pt]{article}

\usepackage{amsmath}
\usepackage{amssymb}
\usepackage{amsfonts}
\usepackage{latexsym}
\usepackage{color}
\usepackage{bm}

\normalbaselines
\catcode `\@=11
\@addtoreset{equation}{section}

\catcode`\@=12

\newtheorem{theo}{Theorem}[section]

\newtheorem{lem}[theo]{Lemma}
\newtheorem{prop}[theo]{Proposition}
\newtheorem{cor}[theo]{Corollary}
\newtheorem{lemma}[theo]{Lemma}

\newtheorem{definition}[theo]{Definition}

\newtheorem{example}[theo]{Example}
 
 \numberwithin{equation}{section}

\newtheorem{remark}[theo]{Remark}

\newcommand{\betheo}{\begin{theo}$\!\!\!${\bf } }
\newcommand{\entheo}{\end{theo}}

\newcommand{\becor}{\begin{cor}$\!\!\!$  }
\newcommand{\encor}{\end{cor}}

\newcommand{\belem}{\begin{lem}$\!\!\!$  }
\newcommand{\enlem}{\end{lem}}

\newcommand{\beprop}{\begin{prop}} 
\newcommand{\enprop}{\end{prop}}

\newcommand{\bedefi}{\begin{definition}$\!\!\!$ \rm }
\newcommand{\findefi}{ \end{definition}}

\newcommand{\beex}{\begin{example}$\!\!\!$ \rm }
\newcommand{\enex}{ \end{example}}

\newcommand{\berem}{\begin{remark}$\!\!\!$ \rm }
\newcommand{\enrem}{ \end{remark}}

\newcommand{\qed}{\hfill $\square$}

\newcommand{\be}{\begin{equation}}
\newcommand{\en}{\end{equation}}

\newcommand{\bea}{\begin{eqnarray}}
\newcommand{\ena}{\end{eqnarray}}

\newcommand{\beano}{\begin{eqnarray*}}
\newcommand{\enano}{\end{eqnarray*}}

\newcommand{\bee}{\begin{enumerate}}
\newcommand{\ene}{\end{enumerate}}

\newcommand{\bei}{\begin{itemize}}
\newcommand{\eni}{\end{itemize}}

\newcommand{\betab}{\begin{tabular}}
\newcommand{\entab}{\end{tabular}}

\newcommand{\up}{\raisebox{0.7mm}{$\upharpoonright $}}%

\newcommand{\subn}[1]{_{\scriptscriptstyle #1}}

\newcommand{\ov}[1]{\overline{#1}}

\newcommand{\mc}{\mathcal}
\newcommand{\mb}{\mathbb}
\newcommand{\RN}{\mb R}

\def\NN{{\mathbb N}}
\def\ZN{{\mathbb Z}}

\def\t{{\mathfrak t}}
\def\D{{\mathcal D}}
\def\F{{\mathcal F}}
\def\H{{\mathcal H}}
\def\I{{\mathcal I}}
\def\J{{\mathcal J}}
\def\K{{\mathcal K}}
\def\M{{\mathcal M}}
\def\O{{\mathcal O}}
\def\P{{\mathcal P}}
\def\R{{\mathcal R}}
\def\T{{\mathcal T}}
\newcommand{\norm}[2]{\left\| #2 \right\|_{#1}}

\newcommand{\noi}{\noindent}
\newcommand{\ud}{\,\mathrm{d}}
\renewcommand{\leq}{\leqslant}
\renewcommand{\geq}{\geqslant}

\newcommand{\BH}{{\mc B}(\H)}

\def\hs{Hilbert space}

\newcommand{\vp}{\varphi}

\newcommand{\ip}[2]{\left\langle {#1}\left|{#2}\right.\right\rangle}

\def\OL{\relax\ifmmode {\sf L}\else{\textsf L}\fi}
\def\OR{\relax\ifmmode {\sf R}\else{\textsf R}\fi}

\newcommand{\pip}{{\sc pip}-space}
\newcommand{\dashvv}{\dashv \!\! \dashv}
\begin{document}

\begin{flushleft}
{\Large \sc Partial inner product spaces, metric operators \\[2mm]  and generalized hermiticity}\vspace*{7mm}

{\large\sf   Jean-Pierre Antoine $\!^{\rm a}$\footnote{{\it E-mail address}: jean-pierre.antoine@uclouvain.be} and
Camillo Trapani $\!^{\rm b}$\footnote{{\it E-mail address}: camillo.trapani@unipa.it}
}
\\[3mm]
$^{\rm a}$ \emph{\small Institut de Recherche en Math\'ematique et  Physique\\
\hspace*{3mm}Universit\'e catholique de Louvain \\
\hspace*{3mm}B-1348   Louvain-la-Neuve, Belgium}
\\[1mm]
$^{\rm b}$ \emph{\small Dipartimento di Matematica e Informatica, Universit\`a di Palermo\\
\hspace*{3mm}I-90123, Palermo, Italy }
\end{flushleft}

\begin{abstract}
Motivated by the recent developments of pseudo-hermitian quantum mechanics,
we analyze the structure of unbounded metric operators in a \hs. It turns out that such operators generate a canonical lattice of \hs s, that is, the simplest case of a partial inner product space (\pip). Next, we introduce several generalizations of the notion of similarity between operators and explore to what extend they preserve spectral properties. Then we apply some of the previous results to operators on a particular \pip, namely, a scale of \hs s generated by a metric operator. Finally, we  reformulate the notion of  pseudo-hermitian operators in the preceding formalism.
\end{abstract}

\noi PACS numbers:  03.65.-w, 03.65.Ca, 02.30.Sa, 02.30.Tb

\section{Introduction }
\label{sect_intro}

Pseudo-hermitian quantum mechanics (QM) is a recent, unconventional, approach to QM, based on the use of non-hermitian Hamiltonians, whose hermitian character can be restored by changing the ambient \hs, via a so-called metric operator.  Such Hamiltonians are not self-adjoint (this is the proper mathematical term, rather than the physicists' hermitian), but have a real spectrum, usually discrete. Instead they are in general $\P\T$-symmetric, that is, invariant under the joint action of space reflection ($\P$) and complex conjugation ($\T$). Typical examples are  the $\P\T$-symmetric, but non-self-adjoint, Hamiltonians   $H = p^2 +ix^3$ and  $H = p^2 -x^4$. Surprisingly, both of them have a purely discrete spectrum, real and positive. A full analysis of $\P\T$-symmetric Hamiltonians may be found in the review paper of Bender \cite{bender}.The motivation comes  from a number of physical problems, mostly from condensed matter physics, but also from scattering theory (complex scaling), relativistic QM and quantum cosmology, or electromagnetic wave propagation in dielectric media.  One may note also that the whole topic is covered in a series of annual international workshops, called ``Pseudo-Hermitian Hamiltonians in Quantum Physics", starting in 2003, the 11th edition having taken place in Paris in August 2012.

These $\P\T$-symmetric Hamiltonians are usually pseudo-hermitian operators, a term introduced a long time ago by
Dieudonn\'e \cite{dieudonne} for characterizing those bounded operators $A$ which satisfy a relation of the form
$A^\ast G = GA$, where $G$ is a \emph{metric operator}, i.e., a strictly positive self-adjoint operator. This operator $G$ then defines a new metric (hence the name) and a new Hilbert space (sometimes called physical) in which $A$ is symmetric and possesses a self-adjoint extension. For a systematic analysis of pseudo-hermitian QM, we may refer to the review of Mostafazadeh \cite{mosta1}.

Now, in most of the literature, the metric operators are assumed to be bounded. In two recent works, however, unbounded
metric operators are introduced \cite{bag-zno, mosta2} in an effort to put the whole machinery on a sound mathematical basis.
In particular, the Dieudonn\'e  relation implies that the operator $A$ is similar to its adjoint $A^\ast$, in some sense, so that this notion of similarity plays a central role in the theory.

Our aim in the present paper is to explore further the structure of unbounded metric operators. We first notice, in Section \ref{sect_2}, that any such an unbounded $G$ generates a lattice of seven \hs s, with lattice operations $\H_1 \wedge \H_2 := \H_1 \cap \H_2 \, ,   \H_1 \vee \H_2 := \H_1 + \H_2$  (see Fig. \ref{fig:diagram}). This structure is then  extended in Section \ref{sect_4} to families of
metric operators, bounded or not. Such a family, if it contains unbounded operators, defines  a rigged \hs, and the latter in turn generates
 a canonical lattice of \hs s. This is a particular case of a  partial inner product space  (\pip),
a concept described at length in our monograph \cite{pip_book}.

Before that, we turn, in Section \ref{sect_3}, to the notion of similarity between operators induced by a bounded metric operator. Since the standard notion is too restrictive for applications, we introduce several generalizations. The most useful one, called \emph{quasi-similarity}, is relevant  when the metric operator is bounded and invertible, but has an unbounded inverse. The goal here is to study which spectral properties are preserved under such a quasi-similarity relation. Remember, we started from non-self-adjoint operators with a discrete real spectrum.

In  Section  \ref{sect_5}, we apply some of the previous results to operators on a \pip, namely, a scale of \hs s generated by the metric operator $G$. The outcome is that the \pip\ structure improves certain results. And finally, in   Section  \ref{sect_6}, we present a construction, inspired from \cite{mosta2}, but significantly more general. Indeed, instead of requiring that the original pseudo-hermitian Hamiltonian $H$ have a countable family of eigenvectors, we only need to assume that $H$ has a (large) set of analytic vectors.

To conclude, let us fix our notations. The framework in a separable \hs\ $\H$, with inner product
$\ip{\cdot}\cdot$, linear in the first entry. Then, for any operator $A$ in   $\H$, we denote its domain by $D(A)$, its range by $R(A)$ and, if $A$ is positive,  its form domain by $Q(A):= D(A^{1/2})$.

\section{Metric operators}
\label{sect_2}

\bedefi A metric operator in a \hs\ $\H$ is a strictly positive self-adjoint operator $G$, that is, $G>0$ or $\ip{G\xi}{\xi}\geq 0$ for every $\xi \in D(G)$ and $\ip{G\xi}{\xi}= 0$ if and only if $\xi=0$.
\findefi
Of course,  $G$ is densely defined and invertible, but need not be bounded; its inverse $G^{-1}$ is also a metric operator,  bounded or not (in this case, in fact, $0\in \sigma_c(G)$).
\berem Let $G, G_1$ and $G_2$ be metric operators. Then
\bei
\item[(1)] If $G_1$ and $G_2$ are both bounded, then $G_1+G_2$ is a bounded metric operator;
\item[(2)] $\lambda G $ is a metric operator if $\lambda>0$;
\item[(3)]  $G^{1/2}$  {and, more generally, $G^{\alpha/2} (-1\leq \alpha \leq 1)$,} is a metric operator.
\eni
\enrem

 \begin{figure}[t]
\centering \setlength{\unitlength}{0.5cm}
\begin{picture}(8,8)

\put(4.2,4){
\begin{picture}(8,8) \thicklines
 \put(-3.4,-0.9){\vector(3,1){2.2}}
\put(-3.4,2.2){\vector(3,1){2}}
 \put(-3.4,-1.9){\vector(3,-1){2.2}}
\put(-3.4,1.2){\vector(3,-1){2.2}}
\put(1.3,0.4){\vector(3,1){2.2}}
\put(1.3,-2.8){\vector(3,1){2.2}}
 \put(1.3,-0.4){\vector(3,-1){2.2}}
\put(1.45,2.6){\vector(3,-1){2.15}}
\put(0,3.2){\makebox(0,0){ $ \H(G^{-1})$}}
\put(-0.1,0){\makebox(0,0){ $\H$}}
\put(0,-3.2){\makebox(0,0){ $\H(G)$}}

\put(-5.3,1.5){\makebox(0,0){ $\H(R_{G^{-1}})$}}
\put(-5.2,-1.5){\makebox(0,0){$\H(R_{G})$}}
\put(5.3,1.5){\makebox(0,0){ $ \H(R_G^{-1})$}}
\put(5.3,-1.5){\makebox(0,0){$ \H(R_{G^{-1}}^{-1})$}}

\end{picture}
}
\end{picture}
\caption{\label{fig:diagram}The lattice of \hs s generated by a metric operator.}

\end{figure}

Now we perform the construction described in \cite[Sec. 5.5]{pip_book}, and largely inspired by interpolation theory \cite{berghlof}.
If $G$ is a metric operator, we consider the domain $D( G^{1/2})$. Equipped with the graph norm,
$$
\norm{R_G}{\xi}^2 := \norm{}{\xi}^2 +\norm{} {G^{1/2}\xi}^2,
$$
this is a \hs, denoted $\H(R_G)$, dense in $\H$. Now we equip that space with the norm
$\norm{G}{\xi}^2 := \norm{}{G^{1/2}\xi}^2$ and denote by $\H(G)$ the completion of $\H(R_G)$ in that norm. Then we have  $\H(R_G) = \H \cap \H(G)$, with the so-called projective norm.
Since $D( G^{1/2}) = Q(G)$, the form domain of $G$, we may also write
$$
\norm{R_G}{\xi}^2  
= \ip{(1+G)\xi}{\xi} = \ip{R_G \xi}{\xi}, \quad \norm{G}{\xi}^2 = \ip{G\xi}{\xi},
$$
where we have put $R_G = 1+G$, which justifies the notation $\H(R_G)$. It follows that the conjugate dual  $\H(R_G)^\times$ of  $\H(R_G)$ is $\H(R_G^{-1})$ and one gets the triplet
\be  \label{eqtr1}
\H(R_G) \;\subset\; \H   \;\subset\;   \H(R_G^{\!-1}).
\en
Proceeding in the same way with the inverse operator $G^{-1}$, we obtain another \hs, $\H(G^{-1})$, and another triplet
\be \label{eqtr2}
\H(R_{G^{-1}})  \;\subset\;  \H   \;\subset\;  \H(R_{G^{-1}}^{-1}).
\en
Then, taking conjugate duals,  it is easy to see that one has
\begin{align}
\H(R_G)^\times &= \H(R_G^{-1}) = \H + \H(G^{-1}),  \label{cup1}\\ 
\H(R_{G^{-1})})^\times &= \H(R_{G^{-1}}^{-1}) = \H + \H(G).   \label{cup2}
\end{align}
Now, if $G$ is bounded, the triplet \eqref{eqtr1} collapses, in the sense that all three spaces coincide as vector spaces, with equivalent norms. Similarly, one gets $\H(R_{G^{-1}}) = \H(G^{-1})$ and
$\H(R_{G^{-1}}^{-1}) = \H(G)$. So we are left with the triplet
\be
  \H(G^{-1}) \;\subset\; \H \;\subset\;  \H(G).
\label{eq:triplet}
\en
If $G^{-1}$ is also bounded, then the spaces $\H(G^{-1})$ and $\H(G)$ coincide with $\H$ and their norms are equivalent to (but different from) the norm of $\H$.

Putting everything together, we get the  diagram shown in Fig. \ref{fig:diagram}.
Note that here  every embedding is continuous and has dense range.

Before proceeding, let us give two (easy) examples, in which $G$ and $G^{-1}$ are multiplication operators in
$\H = L^2(\RN,\ud x)$,  both unbounded, so that the three middle spaces are mutually noncomparable.
\begin{enumerate}
\item The first example comes from  \cite[Sec. 5.5.1]{pip_book}, namely, $G = x^2$, so that $R_G = 1 + x^2$.
Then all spaces appearing in Fig. \ref{fig:diagram} are weighted $L^2$ spaces, as shown in Fig. \ref{fig:diagram2}.
One can check that the norms on the l.h.s. are equivalent to the corresponding projective norms, whereas
those on the r.h.s. are equivalent to the corresponding inductive norms (see the proof of a similar statement for sequences in \cite[Sec. 4.3.1]{pip_book}).

 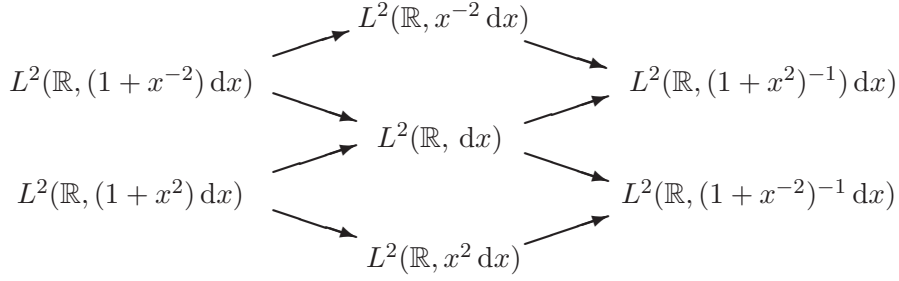
\begin{figure}[t]
\centering \setlength{\unitlength}{0.5cm}
\begin{picture}(8,8)

\put(4.2,4){
\begin{picture}(8,8) \thicklines
 \put(-4.4,-0.9){\vector(3,1){2.2}}
\put(-4.4,2.2){\vector(3,1){2}}
 \put(-4.4,-1.9){\vector(3,-1){2.2}}
\put(-4.4,1.2){\vector(3,-1){2.2}}
\put(2.3,0.4){\vector(3,1){2.2}}
\put(2.3,-2.8){\vector(3,1){2.2}}
 \put(2.3,-0.4){\vector(3,-1){2.2}}
\put(2.3,2.6){\vector(3,-1){2.15}}
\put(0,3.2){\makebox(0,0){ $L^ 2(\RN,x^{-2} \ud x)$}}
\put(-0.1,0){\makebox(0,0){ $L^ 2(\RN, \ud x)$}}
\put(0,-3.2){\makebox(0,0){ $L^ 2(\RN, x^2 \ud x)$}}

\put(-8.3,1.5){\makebox(0,0){ $L^ 2(\RN,(1 + x^{-2})\ud x)$}}
\put(-8.2,-1.5){\makebox(0,0){$ L^ 2(\RN,(1 + x^2)\ud x)$}}
\put(8.5,1.5){\makebox(0,0){ $ L^ 2(\RN,(1 + x^2)^{-1})\ud x)$}}
\put(8.5,-1.5){\makebox(0,0){$L^ 2(\RN,(1 + x^{-2})^{-1}\ud x)$}}

\end{picture}
}
\end{picture}
\caption{\label{fig:diagram2}The lattice of \hs s generated by $G=x^2$.}

\end{figure}
\item For the second example, inspired from \cite{ali-bag-gaz}, one chooses $G = e^{ax}, G^{-1} = e^{-ax}$ and proceeds in the same way (see also the discussion in \cite[Sec. 4.6.3]{pip_book}).
\end{enumerate}

Next, on the space $\H(R_{G})$, the operator $G^{1/2}$ is isometric onto $\H$, hence it extends to a unitary operator from  $\H(G)$ onto $\H$. Analogously, $G^{-1/2}$ is a
unitary operator from  $ \H(G^{-1})$ onto $\H$. In the same way, the operator $R_{G}^{1/2}$ is unitary from $\H(R_{G})$ onto $\H$, and from  $\H$ onto
 $\H(R_G^{-1})$. Hence $R_{G}$ is the Riesz unitary operator mapping $\H(R_{G})$ onto its conjugate dual  $\H(R_G^{-1})$, and similarly $R_{G}^{-1}$ from $\H(R_G^{-1})$ onto
 $\H(R_{G})$, that is, in the triplet \eqref{eqtr1}. Analogous relations hold for $G^{-1}$, i.e. in the triplet \eqref{eqtr2}.

\begin{figure}[h]
\centering \setlength{\unitlength}{0.5cm}
\begin{picture}(8,8)

\put(4.2,4){
\begin{picture}(8,8) \thicklines
 \put(-3.4,-0.9){\vector(3,1){2.2}}
\put(-3.4,2.2){\vector(3,1){2}}
 \put(-3.4,-1.9){\vector(3,-1){2.2}}
\put(-3.4,1.2){\vector(3,-1){2.2}}
\put(1.3,0.4){\vector(3,1){2.2}}
\put(1.3,-2.8){\vector(3,1){2.2}}
 \put(1.3,-0.4){\vector(3,-1){2.2}}
\put(1.45,2.6){\vector(3,-1){2.15}}
\put(0,3.2){\makebox(0,0){ $ G^{-1}$}}
\put(-0.1,0){\makebox(0,0){ $I$}}
\put(0,-3.2){\makebox(0,0){ $G$}}

\put(-5.3,1.5){\makebox(0,0){ $I\wedge G^{-1}$}}
\put(-5.2,-1.5){\makebox(0,0){$I\wedge G$}}
\put(5.3,1.5){\makebox(0,0){ $I \vee G^{-1}$}}
\put(5.3,-1.5){\makebox(0,0){$I\vee G$}}

\end{picture}
}
\end{picture}
\caption{\label{fig:diagram3}The lattice generated by a metric operator.}

\end{figure}
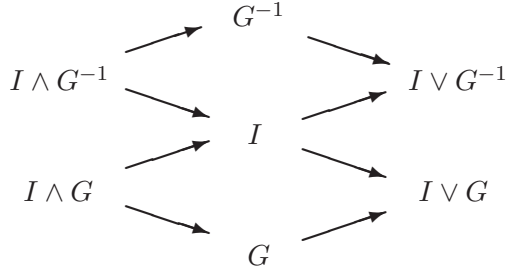

 By the definition of the spaces on the left and the relations \eqref{cup1}-\eqref{cup2},
 it is clear that the spaces on the diagram shown on Fig.  \ref{fig:diagram} constitute a lattice with respect to the lattice operations
 \begin{align*}
 \H_1 \wedge \H_2& := \H_1 \cap \H_2 \, , \\
 \H_1 \vee \H_2& := \H_1 + \H_2 \, .
  \end{align*}
  Since all spaces $\H(A)$ are indexed by the corresponding operator $A$,
  we can as well apply the lattice operations on the operators themselves.
  This would give the diagram shown in Fig. \ref{fig:diagram3}.

The link between the two lattices is given in terms of an order relation:
 $G_1\preceq G_2$ if and only if
 $\H(G_1)\subset \H(G_2)$, where the embedding is continuous and has dense range.
In particular, if $G$ is bounded and $G^{-1}$  unbounded, the relation \eqref{eq:triplet} becomes
$$
G^{-1} \preceq I \preceq G.
$$
In Section \ref{sect_4}, we will extend these considerations to families of metric operators.

Actually one can go further, following a construction made in \cite{scales}. Let $G$ be \emph{unbounded}. Then, if $G > 1$, the norm $\norm{G}{\cdot}$ is equivalent to the graph norm on $D(G^{1/2})$, so that $\H(G) = \H(R_{G})$ and thus also $\H(G^{-1}) = \H(R_{G}^{-1})$.
Hence we get the triplet
\be\label{eq:tri>1}
\H(G) \; \subset\; \H \; \subset\; \H(G^{-1}).
\en
Otherwise, $R_{G} = 1+G > 1$ and it is also a metric operator. Thus we have now
\be\label{eq:tri<1}
\H(R_{G}) \; \subset\; \H \; \subset\; \H(R_{G}^{-1}).
\en
In both cases
one recognizes   that the triplet \eqref{eq:tri>1}, resp. \eqref{eq:tri<1}, is the central part of the discrete scale of Hilbert spaces built on the powers of $G^{1/2}$, resp. $R_{G}^{1/2}$.
This means, in the first case, $V_{\J}:= \{\H_{n}, n \in \ZN \}$,
where $\H_{n} = D(G^{n/2}),  n\in \NN$, with a norm equivalent to the graph norm, and $ \H_{-n} =\H_{n}^\times$:
\be\label{eq:scale}
 \ldots\subset\; \H_{2}\; \subset\;\H_{1}\; \subset\; \H \; \subset\; \H_{-1} \subset\; \H_{-2} \subset\; \ldots
\en
Thus $\H_{1} =  \H(G )$ and $\H_{-1} =  \H(G^{-1})$.
In the second case, one simply replaces $G^{1/2}$ by $R_{G}^{1/2}$.

As in the original construction, this raises the question of identifying the end spaces of the scale, namely,
\be \label{eq:endscale}
\H_{\infty}(G^{1/2}):=\bigcap_{n\in \ZN} \H_n, \qquad \H_{-\infty}(G^{1/2}):=\bigcup_{n\in \ZN} \H_{n}.
\end{equation}
In fact, one can go one more step. Namely, following \cite[Sec. 5.1.2]{pip_book}, we can use quadratic interpolation theory and build a continuous scale of Hilbert spaces
$\H_{\alpha}, 0\leq \alpha\leq 1$, between  $\H_{1}$  and $\H $, where $\H_{\alpha}=  D(G^{\alpha/2})$, with norm $\|\xi\|_{\alpha} = \|G^{\alpha/2}\xi\|$.
Notice that every $G^\alpha, \alpha\geq 0$, is a bounded metric operator.

Next we define $\H_{-\alpha} =\H_{\alpha}^\times$ and iterate the construction to the full continuous scale $V_{\widetilde \J}:= \{\H_{\alpha}, \alpha \in \RN \}$.
Then, of course, one can replace $\ZN$ by $\RN$ in the definition \eqref{eq:endscale} of the end spaces of the scale.

Let us give two (trivial) examples. Take first $\H = L^2(\RN, \ud x)$ and define
$G_x$ as the operator of multiplication  by $(1+x^2)^{1/2}$, which is an unbounded metric operator. In the same way, define $G_p := (1-\ud^2\!/\!\ud x^2)^{1/2} = \F G_x \F{-1}$, where $\F$ is the Fourier transform. Similarly, in $L^2(\RN^3)$, we can take $G_p := (1-\Delta)^{1/2}$.
For these examples, the end spaces  of the scale \eqref{eq:scale} are easy to identify:
$\H_{\infty}(G_x^{1/2})$ consists of square integrable,  fast decreasing functions, whereas
the scale built on $G_p$ is precisely the scale of Sobolev spaces.

More generally, given any unbounded self-adjoint operator $A$ in $\H$, $G_A:= (1+A^2)^{1/2}$ is an unbounded metric operator, larger than 1, and the construction of the corresponding scale is straightforward.

\section{Similar and quasi-similar operators}
\label{sect_3}

In this section we collect some basic definitions and facts about similarity of linear operators in Hilbert spaces and discuss
several generalizations of this notion. Throughout the section, $G$ will denote a \emph{bounded} metric operator.
We begin by the following easy results.
\begin{lemma}\label{lemma_212} Given the bounded metric operator $G$, let $A$ be  a linear operator in $\H$ with $D(A)$ dense in $\H$. Then $D(A)$ is dense in $\H(G)$.
\end{lemma}
 {\bf Proof. } Every $\xi \in \H(G)$ is the limit of a sequence in $\H$. Each element of this sequence is, in turn, the limit of a sequence of elements of $D(A)$,
with respect to the norm of $\H$ which is stronger than the norm of $\H(G)$.
\qed
\begin{prop} \label{prop-16}
Let $A$ be  a linear operator in $\H$. If $A$ is closed in $\H(G)$, then it closed in $\H$.
\end{prop}
 {\bf Proof. } Let $\{\xi_n\}$ be  a sequence in $D(A)$ such that $\xi_n \to \xi$ and $A\xi_n$ is Cauchy in $\H$. Then
$\xi_n \stackrel{G}{\to} \xi$ and $A\xi_n$ is $G$-Cauchy.  Hence $\xi \in D(A)$ and $A\xi_n \stackrel{G}{\to} A\xi$. Then, it follows easily that
$A\xi_n {\to} A\xi$.
\qed
\smallskip

 Notice that, both in Lemma \ref{lemma_212} and in Proposition \ref{prop-16}, one can replace the space $\H(G)$ by $\H(G^\alpha)$, for any $\alpha>0$.

\subsection{Similarity}\label{sect_sim}

Let $\H, \K$ be Hilbert spaces, $D(A)$ and $D(B)$ dense subspaces\footnote{
From now on we will always suppose the domains of the given operators to be dense in $\H$.}
of $\H$ and $\K$,  respectively,   $A:D(A) \to \H$, $B: D(B) \to \K$ two linear operators.
 A bounded operator $T:\H \to \K$ is called an \emph{intertwining operator}  for $A$ and $B$ if
\begin{itemize}
\item[({\sf io$_1$})] $T:D(A)\to D(B)$;
\item[({\sf io$_2$})]$BT\xi = TA\xi, \; \forall\, \xi \in D(A)$.
\end{itemize}

{ \berem \label{rem_adjoint}If $T$ is an intertwining operator for $A$ and $B$, then $T^* :\K \to \H$ is an intertwining operator for $B^*$ and $A^*$.

\enrem}

\bedefi
Let  $A, B$ be two linear operators in the Hilbert spaces $\H$ and $\K$, respectively. Then,

(i) We say that $A$ and $B$ are \emph{similar}, and write $A\sim B$,
if there exists an intertwining operator $T$ for $A$ and $B$ with bounded inverse $T^{-1}:\K\to \H$, intertwining for $B$ and $A$ .

(ii) $A$ and $B$ are \emph{unitarily equivalent} if $A\sim B$ and $T:\H \to \K$ is unitary, in which case we write $A\approx B$.   

\findefi
\berem   We notice that  $\sim$ and $\approx$ are equivalence relations.   Also, in that case, $TD(A) = D(B)$.
\end{remark}

The following properties of similar operators are easily proved.

\begin{prop} \label{prop_closedness} Let $A, B$ be linear operators in $\H$ and $\K$, respectively.
The following statements hold.
\begin{itemize}
\item[(i)]{ $A\sim B$ if, and only if $B^* \sim A^*$.}
\item[(ii)] $A$ is closed if, and only if, $B$ is closed.
\item[(iii)] $A^{-1}$ exists if, and only if,  $B^{-1}$ exists. Moreover, $B^{-1} \sim A^{-1}$.
\end{itemize}
\end{prop}
 {\bf Proof. } We denote by $T$ the intertwining operator for $A$ and $B$.

{ (i) follows from Remark \ref{rem_adjoint}.}

(ii): Assume that $B$ is closed. Let $\{\xi_n\}$ be a sequence in $D(A)$ such that $\xi_n \to \xi$ and $A\xi_n \to \eta$. Then, $T\xi_n \to T\xi$ and $TA\xi_n \to T\eta$. But $TA\xi_n = BT\xi_n\to T\eta$. Hence $T\xi \in D(B)$ and $BT\xi = T\eta$. The assumption implies that $\xi \in T^{-1}(D(B))=D(A)$ and
$TA\xi =T\eta$ or, equivalently, $A\xi=\eta$. Hence, $A$ is closed. The statement follows by replacing $A$ with $B$ (and $T$ with $T^{-1}$).

(iii): Let $\eta \in D(B)$ and $B\eta=0$. By assumption, there exists a unique $\xi \in D(A)$ such that $\eta = T\xi$. Hence,
$ TA\xi= BT\xi= B\eta=0$. This implies $A\xi=0$ and, in turn $\xi=0$. Thus $B$ is invertible. Moreover,
$D(B^{-1})= B(D(B))= BT(T^{-1}D(B))= BT(D(A))=TA(D(A))= T(D(A^{-1}))$. Hence, if $\eta \in D(B^{-1})$, $T^{-1}\eta \in D(A^{-1})$.
The equality $T^{-1}B^{-1}\eta = A^{-1}T^{-1}\eta$, for every $\eta \in D(B^{-1})$ follows easily. The proof is completed by replacing $A$ with $B$. 
\qed

\begin{prop} \label{prop37}
Let $A$, $B$ be closed operators. Assume that $A\sim B$. Then $\rho(A)= \rho(B)$.
\end{prop}
 {\bf Proof. }
Let $\lambda \in \rho(A)$, the resolvent set of $A$, then $(A-\lambda I)^{-1}$ exists and it is bounded. Define $X_\lambda=T(A-\lambda I)^{-1}T^{-1}$.
Then, $X_\lambda$ is bounded. Since $(A-\lambda I)^{-1}T^{-1}\eta \in D(A)$, for every $\eta \in \K$, we have
\begin{align*}
(B-\lambda I)X_\lambda \eta &= (B-\lambda I) T(A-\lambda I)^{-1}T^{-1}\eta \\
&= T(A-\lambda I) (A-\lambda I)^{-1}T^{-1}\eta = \eta , \quad \forall \eta \in \K.
\end{align*}
On the other hand, since $(B-\lambda I)T\xi = T(A-\lambda I)\xi$, for all $\xi \in D(A)$, taking $\xi =T^{-1}\eta$, $\eta \in D(B)$, we obtain
$(B-\lambda I)\eta = T(A-\lambda I)T^{-1}\eta$ and then $T^{-1}(B-\lambda I)\eta= (A-\lambda I)T^{-1}\eta)$.
Then, for every $\eta \in D(B)$ we get
\begin{align*}
 X_\lambda (B-\lambda I)\eta &= T(A-\lambda I)^{-1}T^{-1}(B-\lambda I)\eta\\
 &= T(A-\lambda I)^{-1}(A-\lambda I)T^{-1}\eta = \eta, \quad \forall \eta \in D(B).
\end{align*}
Hence $X_\lambda = (B-\lambda I)^{-1}$ and $\lambda \in \rho(B)$.

The statement follows by replacing $A$ with $B$ (and $T$ with $T^{-1}$).
\qed
\berem In the previous proof both  assumptions `$T^{-1}$ bounded' and `$TD(A)=D(B)$' seem to be unavoidable: the first guarantees that $X_\lambda$
 (which is in any case a left inverse) is bounded; the second allows to prove that $X_\lambda$ is also a right inverse.
\enrem

Similarity of $A$ and $B$ is symmetric, preserves both the closedness of the operators and their spectra. But, in general, it does not preserve self-adjointness.

Similarity preserves also the parts in which the spectrum is traditionally decomposed: the point spectrum $\sigma_p (\cdot)$, the continuous spectrum $\sigma_c(\cdot)$
and the residual spectrum $\sigma_r(\cdot)$.

\begin{prop}\label{prop_spectrum_sim} Let $A$, $B$ be closed operators. Assume that $A\sim B$. Then,
 \begin{itemize} \item[(i)] $\sigma_p(A)=\sigma_p(B)$. Moreover is $\xi \in D(A)$ is an eigenvector of $A$ corresponding to the eigenvalue $\lambda$,
 then $T\xi$ is an eigenvector of $B$ corresponding to the same eigenvalue. Conversely, if $\eta \in D(B)$ is an eigenvector of $B$ corresponding to the eigenvalue $\lambda$,
then $T^{-1}\eta$ is an an eigenvector of $A$ corresponding to the same eigenvalue. Moreover, the multiplicity of $\lambda$ as eigenvalue of $A$ is the same as its multiplicity
 as eigenvalue of $B$.
\item[(ii)] $\sigma_c(A)= \sigma_c(B).$
\item[(iii)]$\sigma_r(A)= \sigma_r(B)$.
\end{itemize}
\end{prop}
 {\bf Proof. } The first statement is very easy. We prove (ii). Let $\lambda \in \sigma_c(B)$.  If $\eta \in \H$, then $\eta= T^{-1}\eta '$ for some $\eta ' \in \K$.
 Since $R(B-\lambda I)$ is dense in $\K$, there exists a sequence $\{\eta'_k \}\subset R(B-\lambda I)$ such that $\eta'_k \to \eta'$. Put $\eta'_k= (B-\lambda I)\xi' _k$,
 with $\xi'_k \in D(B)$. Since $TD(A)=D(B)$, for every $k \in \mb N$ there exists $\xi_k \in D(A)$ such that $ \xi'_k =T\xi_k$. Hence
$$ \eta'= \lim_{k\to \infty} (B-\lambda I)T\xi_k= \lim_{k\to \infty}T(A-\lambda I)\xi_k.$$
This implies that
$$ \eta = T^{-1}\eta'= \lim_{k\to \infty} (A-\lambda I)\xi_k.$$
Thus $R(A-\lambda I)$ is dense in $\H$. The unboundedness of $(B-\lambda I)^{-1}$ implies easily the unboundedness of  $(A-\lambda I)^{-1}$.
 Hence $\sigma_c(A)\subseteq \sigma_c(B).$ Interchanging the roles of $A$ and $B$ one gets the reverse inclusion. (iii) follows from Proposition \ref{prop37} and (i), (ii).
\qed

Taking into account that, if $A$ is self-adjoint, its residual spectrum is empty, we obtain
\becor
Let $A$, $B$ be closed operators with $A$ self-adjoint. Assume that $A\sim B$. Then $B$ has real spectrum and $\sigma_r(B)=\emptyset$.
\encor

This corollary  can be used to show the existence of non symmetric operators having real spectrum and empty residual spectrum.
{On the other hand, under certain conditions, an operator similar to its adjoint (i.e. a \emph{pseudo-hermitian} operator \cite{bender,dieudonne,mosta1}) is automatically self-adjoint \cite{williams}.}

\subsection{Quasi-similitarity and spectra}

The notion of similarity discussed in the previous section is too strong in many situations.
\bedefi
We say that $A$ and $B$ are \emph{quasi-similar}, and write $A\dashv B$, if there exists an intertwining operator $T$ for $A$ and $B$ which is  invertible, with inverse $T^{-1}$   densely defined (but not necessarily bounded).
\findefi

 \berem (1) Even if $T^{-1}$ is bounded, $A$ and $B$ need not be similar, unless $T^{-1}$ also is an intertwining operator.

(2) {According to Dieudonn\'e \cite{dieudonne}, a  \emph{quasi-hermitian} operator is a {bounded} operator $A$ satisfying the relation $A^\ast T = TA$, where $T>0$, but $T^{-1}$ is  not necessarily bounded. The same notion has been introduced  by Sz.-Nagy and  Foia\c{s} \cite[Chap.II, Sec.3]{nagy}, who call it \emph{quasi-affinity}.} For unbounded operators, a slightly more restrictive notion, but under the same name of quasi-similarity, is due to   \^{O}ta and  Schm\"udgen  \cite{ota-schm}.
\enrem

{\begin{prop} \label{prop_adjoint} If $A\dashv B$, with intertwining operator $T$, then $B^*\dashv A^*$ with intertwining operator $T^*$.
\end{prop}
 {\bf Proof. } This follows from Remark \ref{rem_adjoint} and from the fact that, since $T^{-1}$ exists, then ${(T^*)}^{-1}$ exists too and ${(T^*)}^{-1}={(T^{-1})}^*$.
 \qed
 }
\bedefi
The operators $A$ and $B$ are called \emph{weakly quasi-similar}, in which case we write $A\dashv_w B$, if $B$ is closable,
$T$ is invertible  with densely defined inverse $T^{-1}$    and, instead of ({\sf io$_1$}) and ({\sf io$_2$}), the following condition holds
\begin{itemize}
\item[({\sf ws})] $\ip{T\xi}{B^*\eta}= \ip{TA\xi}{\eta}, \; \forall \xi \in D(A),\, \eta \in D(B^*).$
\end{itemize}
\findefi

\begin{prop}
  $A\dashv_w B$  if and only if $T:D(A) \to D(B^{**})$ and $B^{**}T\xi= TA\xi$, for every $\xi \in D(A)$. In particular if $B$ is closed, $A\dashv B$ if,
and only if, $A\dashv_w B$.
\end{prop}

\begin{prop}\label{prop_closable} If $B$ is closable and $A\dashv_w B$, then $A$ is closable.
\end{prop}
 {\bf Proof. }
Assume that $\{\xi_n\}$ is a sequence in $D(A)$ and $\xi_n \to 0$, $A\xi_n \to \eta$. Then, $T\xi_n \to 0$ and $TA\xi_n \to T\eta$.
But $TA\xi_n = B^{**}T\xi_n\to T\eta$. From the closedness of $B^{**}$ it follows that $T\eta =0$ and, therefore, $\eta=0$.
\qed

\beex \label{ex_1} The converse of the previous statement does not hold, in general. For instance, in the Hilbert space $L^2 ({\mb R})$, consider the operator $Q$ defined on the dense domain
$$
 D(Q)= \left\{ f \in L^2 ({\mb R}): \int_{\mb R} x^2 |f(x)|^2 dx < \infty \right\}.
$$
Given $\varphi \in L^2 ({\mb R})$, with $\norm{}{\varphi}=1$,
let $P_\varphi:= \varphi \otimes \overline{ \varphi}$ denote the projection operator onto the one-dimensional subspace generated by $\varphi$ and $A_\varphi$ the operator with domain $D(A_\varphi)= D(Q^2)$ defined by
$$
A_\varphi f = \ip{(I+Q^2)f}{\varphi}(I+Q^2)^{-1}\varphi, \quad \varphi \in D(A_\varphi).
$$
Then, it is easily seen that $P_\varphi \dashv A_\varphi$ with the intertwining operator $T:= $ \mbox{$(I+Q^2)^{-1}$}. Clearly $P_\varphi$ is everywhere defined and bounded, but the operator $A_\varphi$ is closable if, and only if, $\varphi \in D(Q^2)$.
\enex

\bedefi If $A\sim B$ (resp., $A \dashv B$) and the intertwining operator is a metric operator $G$, we say that $A$ and $B$ are {\em metrically} similar (resp., quasi-similar).
\findefi

If $A\sim B$ and $T$ is the corresponding intertwining operator, then $T=UG$, where $U$ is unitary and $G:=(T^*T)^{1/2}$ is a metric operator.
If we put $B'= U^{-1}BU$, then $B'$ and $A$ are metrically similar. Thus, up to unitary equivalence, one can always consider metric similarity instead of similarity.
\medskip

In the sequel of this section, we take a fixed metric operator $G$.

\begin{prop}\label{prop_214}Let  $A$ be  a closed  operator in $\H$.
Put
$$
 D(A^\star_G):= \{\eta \in \H: G\eta \in D(A^*), \, A^*G\eta \in D(G^{-1}) \}
$$
and $$
A^\star_G \eta := G^{-1}A^*G\eta, \quad \forall \eta \in D(A^\star_G).
$$
Then $A^\star_G$ is the restriction to $\H$ of the adjoint $A^*_G$ of $A$ in $\H(G)$.
\end{prop}
 {\bf Proof. } Let $\xi \in D(A)$ and $\eta \in D(A^\star_G)$. Then,
\begin{align*}
\ip{A\xi}{\eta}_G &= \ip{GA\xi}{\eta} =\ip{A\xi}{G\eta}\\
&= \ip{\xi}{A^*G\eta}=\ip{G^{-1}G\xi}{A^*G\eta}\\
 &=\ip{G\xi}{G^{-1}A^*G\eta}=\ip{\xi}{G^{-1}A^*G\eta}_G.\\[-14mm]
&
\end{align*}
\qed
\smallskip

\berem $D(A^\star_G)$ need not be dense in $\H$. We do not know whether $A$ is closed or closable in $\H(G)$; so the existence of a nontrivial adjoint is not guaranteed.
\enrem

\begin{prop}\label{prop_216}Let  $A$ be  a closed operator in $\H$.
 If the subspace
 $$
D(A^\star_G):= \{\eta \in \H: G\eta \in D(A^*), \, A^*G\eta \in D(G^{-1}) \}
$$
 is dense in $\H$, then $A$ has a densely defined adjoint $A^*_G$ in $\H(G)$ and $A$ is closable in $\H(G)$.
\end{prop}
\berem We do not know if $(A^*_G)^*_G = A$, i.e. if $A$ is  closed in $\H(G)$.
\enrem

 Let $A,B$ be closed  operators in $\H$. Assume that they are  {metrically quasi-similar} and let $G$ be the metric intertwining operator
for $A,B$.
Then, by {\sf (ws)} we have
$$ \ip{G\xi}{B^*\eta}=\ip{GA\xi}{\eta}, \quad \forall \xi \in D(A), \, \eta \in D(B^*).$$
This equality can be rewritten as
$$ \ip{\xi}{B^*\eta}_G=\ip{A\xi}{\eta}_G, \quad \forall \xi \in D(A), \, \eta \in D(B^*).$$
This means that $B^*$ is a restriction of $A^*_G$, the adjoint of $A$ in $\H(G)$.

Since $D(B^*)$ is dense in $\H$, then, by Lemma \ref{lemma_212}, $A^*_G$ is densely defined in $\H(G)$ and $(A^*_G)^*_G \supseteq A$.
Then we can consider, as in Proposition \ref{prop_216}, the operator $A^\star_G$. This operator is an extension of $B^*$ in $\H$.

Clearly $A^\star_G$ satisfies the equality
$$ \ip{\xi}{A^\star_G\eta}_G=\ip{A\xi}{\eta}_G, \quad \forall \xi \in D(A), \, \eta \in D(A^\star_G)$$
and by Proposition \ref{prop_214}, $A^\star_G \eta := G^{-1}A^*G\eta, \quad \forall \eta \in D(A^\star_G)$. Since $A^\star_G\supset B^*$, then $(A^\star_G)^* \subset B$.
Put $B_0:= (A^\star_G)^*$. Then $B_0$ is minimal among the closed operators $B$ satisfying, for fixed $A$ and $G$, the conditions
\begin{align*} & G:D(A)\to D(B);\\
& BG\xi = GA\xi, \; \forall \xi \in D(A).
\end{align*}
From these facts it follows easily that
$GD(A)$ is a core for $B_0$. Indeed, it is easily checked that $GD(A)$ is dense in $\H$. Let $B_1$ denote the closure of the restriction of $B_0$ to $GD(A)$.
Then $B_1\subseteq B_0$ and it is easily seen that $B_1$ satisfies the two conditions above. Hence $B_1=B_0$.

 {Thus we have proved:
\belem Let $A,B$ be closed and  $A\dashv B$ with a metric intertwining operator $G$. Then
$A^\star_G$ is densely defined, $B_0:= (A^\star_G)^*$  is minimal among the closed operators $B$ satisfying, for fixed $A$ and $G$, the conditions
\begin{align*} & G:D(A)\to D(B);\\
& BG\xi = GA\xi, \; \forall \xi \in D(A),
\end{align*}
and $GD(A)$ is a core for $B_0$.
\enlem}
\medskip
Now we consider the relationship between the spectra of quasi-similar operators.

\begin{prop}\label{prop_sigmap} Let $A$ and $B$ be closed operators and assume that $A\dashv B$,  with intertwining operator $T$. The following statements hold
\begin{itemize}
\item[(i)] $\sigma_p(A)\subseteq \sigma_p(B)$ and for every $\lambda \in \sigma_p(A)$ one has $m_A(\lambda) \leq m_B(\lambda)$, where $m_A(\lambda)$, resp. $ m_B(\lambda)$, denotes  the multiplicity of $\lambda$ as eigenvalue of the operator $A$, resp. $B$.

\item[(ii)]  $\sigma_r(B) \subseteq\sigma_r(A)$.

\item[(iii)] If $TD(A)=D(B)$, then $\sigma_p(B)= \sigma_p(A)$. 
\item[(iv)] If $T^{-1}$ is bounded and $TD(A)$ is a core for B,  then $\sigma_p(B)\subseteq \sigma(A)$.
\end{itemize}
\end{prop}

 {\bf Proof. }

{ The statements (i) and (iii) can be proved as in Proposition \ref{prop_spectrum_sim}. We prove only the  statements (ii) and (iv).

(ii) By Proposition \ref{prop_adjoint}, $B^*\dashv A^*$, with intertwining operator $T^*$. Then, by (i), $\sigma_p(B^*)\subseteq \sigma_p (A^*)$. The statement follows by observing that $\sigma_r(C)= \overline{\sigma_p(C^*)}=\{ \overline{\lambda}:\lambda \in \sigma_p(C^*)\}$, for every closed operator $C$.  }

(iv): Let $\lambda \in \sigma_p(B)$. Then there exists $\eta \in D(B)\setminus\{0\}$ such that $B\eta= \lambda \eta$.
We may suppose that $\|\eta\|=1$.
Since  $TD(A)$ is a core for $B$, there exists a sequence $\{\xi_n\}\subset D(A)$ such that $T\xi_n \to \eta$ and $BT\xi_n \to B\eta$. Then,
\begin{align*}
\lim_{n \to \infty} T(A\xi_n -\lambda \xi_n)&= \lim_{n \to \infty} TA\xi_n -\lambda \lim_{n \to \infty}T\xi_n=\lim_{n \to \infty}BT\xi_n -\lambda \eta\\
&= B\eta -\lambda \eta =0.
\end{align*}
By the boundedness of $T^{-1}$,
\begin{equation}\label{eqn_non} \lim_{n \to \infty}(A\xi_n -\lambda \xi_n)=0,\end{equation}
Assume that $\lambda \in \rho(A)$. Then $(A-\lambda I)^{-1} \in \BH$. We put, $\eta_{n}= (A-\lambda I)\xi_{n}$.
Then, by \eqref{eqn_non}, $\eta_{n}{\to} 0$. Hence, $\xi_{n}= (A-\lambda I)^{-1}\eta_{n}{\to} 0$. This in turn implies
that $T\xi_{n} {\to} 0$, which is impossible since $\|\eta\|=1$.
\qed

\begin{prop}\label{lemma_one} Let $A$ and $B$ be closed operators. Assume that $A\dashv B$, with intertwining operator $T$.
Then the following statements hold.
\begin{itemize}
\item[(a)] Let $\lambda \in \rho(A)$ and define
\begin{align*} D(X_\lambda) &=D(T^{-1})\\
  X_\lambda\eta &=T(A-\lambda I)^{-1}T^{-1}\eta, \; \eta \in D(X_\lambda).
\end{align*}
Then,
\begin{itemize}\item[(a.1)] $(B-\lambda I) X_\lambda \eta = \eta, \; \forall \eta \in D(X_\lambda)$.
\vspace*{1mm}
\item[(a.2)] If $ (B-\lambda I)\eta \in D(T^{-1}), \; \forall \eta \in D(B)$, and $\lambda \not\in \sigma_p(B)$, then
$ X_\lambda (B-\lambda I)\eta = \eta ,\; \forall \eta \in D(B)$.
\end{itemize}
\vspace*{1mm}
\item[(b)] Let $\lambda \in \rho(B)$ and define
\begin{align*}  D(Y_\lambda) &=\{\xi \in \H:\, (B-\lambda I)^{-1}T\xi \in D(T^{-1})\}\\
  Y_\lambda\xi &=T^{-1}(B-\lambda I)^{-1}T\xi, \; \xi \in D(Y_\lambda).
\end{align*}
Then,
\begin{itemize}\item[(b.1)]
$Y_\lambda(A-\lambda I)\xi =\xi,\; \forall \xi \in D(A)$.
\vspace*{1mm}
\item[(b.2)] For every $\eta \in \H$ such that $Y_\lambda\eta \in D(A)$, $(A-\lambda I)Y_\lambda\eta=\eta.$
\end{itemize}
\end{itemize}
\end{prop}
 {\bf Proof. } (a.1): Let $\lambda \in \rho(A)$, then $(A-\lambda I)^{-1}$ exists in $\BH$. Since \\
\mbox{$(A-\lambda I)^{-1}T^{-1}\eta \in D(A)$},
for every $\eta \in  D(T^{-1})$, we have
\begin{align}\label{eqn_inv}
(B-\lambda I)X_\lambda \eta &= (B-\lambda I) T(A-\lambda I)^{-1}T^{-1}\eta \\
&= T(A-\lambda I) (A-\lambda I)^{-1}T^{-1}\eta = \eta. \nonumber
\end{align}

(a.2): Assume that $(B-\lambda I)\eta \in D(T^{-1}), \; \forall \eta \in D(B)$. Then
$$\
\eta':=T(A-\lambda I)^{-1}T^{-1}(B-\lambda I)\eta
$$
  is well defined.
Observing that $T^{-1} \eta' \in D(A)$, we have
$$
T(A-\lambda I)T^{-1} \eta'= (B-\lambda I)\eta.
$$
Hence, using the quasi-similarity,
$$
(B-\lambda I)TT^{-1} \eta' =(B-\lambda I)\eta.
$$
Thus  $(B-\lambda I)(\eta- \eta')=0$. If $\lambda \not\in \sigma_p(B)$, it follows that $\eta=\eta'$.
This implies that
$$
X_\lambda (B-\lambda I)\eta = \eta ,\; \forall \eta \in D(B).
$$
(b.1): If $\xi \in D(A)$, we have
$$ (B-\lambda I)T\xi = T(A-\lambda I)\xi,$$
i.e.,
$$
T\xi= (B-\lambda I)^{-1}T(A-\lambda I)\xi.
$$
Hence, $(B-\lambda I)^{-1}T(A-\lambda I)\xi \in D(T^{-1})$ and, thus,
$$
\xi = Y_\lambda (A-\lambda I)\xi, \; \forall \xi \in D(A).
$$
(b.2): Let  $\eta \in \H$ with $Y_\lambda\eta \in D(A)$. Put $(A-\lambda I)Y_\lambda \eta := \eta'$. Then, using the quasi-similarity, we have
$$
 (B-\lambda I)TY_\lambda \eta= T(A-\lambda I)Y_\lambda \eta= T\eta'.
$$
The l.h.s. equals $T\eta$. Hence $T\eta=T\eta'$ and so $\eta=\eta'$. In conclusion,
$$
 (A-\lambda I)Y_\lambda\eta=\eta, \quad \forall \eta:\, Y_\lambda\eta \in D(A).
$$
\qed

\becor \label{cor_3.24} Let $A$, $B$ be as in Proposition \ref{lemma_one} and assume that $T^{-1}$ is everywhere defined and bounded. Then $$\rho(A)\setminus \sigma_p(B) \subseteq \rho(B),$$
and
$$ \rho(B)\setminus \sigma_r(A) \subseteq \rho(A).$$
\encor
 {\bf Proof. }
The first inclusion  is an immediate application of (a) of the previous proposition and the second is obtained by taking the adjoints. \makebox[3cm]{}
\qed

\becor \label{cor_two} Let $A,B$ be as in Proposition \ref{lemma_one}. Assume that $D(B)$ and  $ R(B)$ are subspaces of $D(T^{-1})$. Then
$\rho(A)\setminus \sigma_p(B)  \subseteq \rho(B)\cup  \sigma_c(B)$.
\encor
 {{\bf Proof. } Let $\lambda \in \rho(A)\setminus \sigma_p(B)$. By Proposition \ref{lemma_one}(a), the operator $(B-\lambda I)^{-1}$ has a densely defined inverse. If
$(B-\lambda I)^{-1}$ is bounded, then it has an everywhere defined bounded closure,
which coincides with $(B-\lambda I)^{-1}$, since the latter is closed, being the inverse of a closed operator. In this case, $\lambda \in \rho(B)$. If
$(B-\lambda I)^{-1}$
is unbounded, then $\lambda \in \sigma_c(B)$. In other words, $\rho(A)\setminus \sigma_p(B)  \subseteq \rho(B)\cup  \sigma_c(B)$.
\qed
\smallskip

  {Let us consider again the special case where $T^{-1}$ is also everywhere defined and bounded (but does not necessarily satisfy $TD(A)=D(B)$).

\beprop \label{prop_220}
Let $A,B$ be as in Proposition \ref{lemma_one}. Assume that $T^{-1}$ is everywhere defined and bounded and $TD(A)$ is a core for $B$. Then
$$
 \sigma_p (A)\subseteq \sigma_p (B) \subseteq \sigma (B) \subseteq \sigma(A).
$$
\enprop
 {\bf Proof. } We simply notice that, in this case, by (ii) of Proposition \ref{prop_sigmap}, $\sigma_p(B)\subset \sigma(A)$. Hence,
$\rho(A)\setminus \sigma_p(B) =\rho(A) \subseteq \rho(B)$, by Corollary \ref{cor_3.24}.
\qed
 }

  {\berem The situation described in Proposition \ref{prop_220} is quite important for possible applications.
Even if the spectra of $A$ and $B$ may be different, it gives a certain number of informations on $\sigma(B)$
once $\sigma(A)$ is known.  For instance, if $A$ has a pure point spectrum, then $B$ is isospectral to $A$.
More generally, if $A$ is self-adjoint, then any operator $B$ which is quasi-similar to $A$ by
means of an intertwining operator $T$ whose inverse is bounded too, has real spectrum.

\enrem }
\beex Let us consider the operators $P_\vp$ and $A_\vp$ of Example \ref{ex_1} with $\vp \in D(Q^2)$. In this case $A_\vp$ is bounded and everywhere defined and, as noticed before, $P_\varphi \dashv A_\varphi$ with the intertwining operator $T:= $ \mbox{$(I+Q^2)^{-1}$} The spectrum of $A_\vp$ is easily computed to be $\sigma(A_\vp)=\{0, 1\}$. Thus it coincides with $\sigma(P_\vp)$. To see this, we begin by  looking for eigenvalues. The equation
\begin{equation}\label{eq_eigenvalues}
\ip{(I+Q^2)f}{\varphi}(I+Q^2)^{-1}\varphi -\lambda f=0
\end{equation}
has non zero solutions in two cases: if $\lambda=0$, then any element of $\{(I+Q^2)\vp\}^\perp$ is an eigenvector. If $\lambda \neq 0$, then a solution must be a multiple of $(I+Q^2)^{-1}\varphi $, i.e., $f= \kappa (I+Q^2)^{-1}\varphi$. Substituting in \eqref{eq_eigenvalues} one obtains $\lambda =1$ and the set of eigenvectors is the one-dimensional subspace generated by
$(I+Q^2)^{-1}\varphi$. On the other hand, if $\lambda \not\in \{0,1\}$, then, for every $g \in L^2({\mb R})$, the equation
$(A_\vp-\lambda I)f= g$ has the unique solution
$$
f= - \frac{1}{\lambda}g+ \frac{\ip{g}{(I+Q^2)\vp}}{\lambda(1-\lambda)}(I+Q^2)^{-1}\vp. 
$$
Thus, $(A_\vp-\lambda I)^{-1}$ is an everywhere defined bounded operator.
We then conclude that $\sigma(A_\vp)=\sigma(P_\vp)=\{0, 1\}$.
\enex
\beex  Let $A$ be the operator in $L^2({\mb R})$ defined as follows:
\begin{align*}
& D(A)= W^{1,2}({\mb R}),  \\
& (Af)(x)=f'(x)- \frac{2x}{1+x^2}f(x), \quad f\in D(A).
\end{align*}
Then $A$ is a closed operator in $L^2({\mb R})$, being the sum of a closed operator and a bounded one. Let $B$ be the closed operator defined by
 \begin{align*}
& D(B)= W^{1,2}({\mb R}),\\
& (Bf)(x)=f'(x), \quad f \in D(B).
\end{align*}

Then $A\dashv B$ with the intertwining operator $T=(I+Q^2)^{-1}$. Indeed, it is easily seen that $T: W^{1,2}({\mb R})\to W^{1,2}({\mb R})$. Moreover, for every $f \in W^{1,2}({\mb R})$, one has
$$
(TAf)(x)= (1+x^2)^{-1} \left(f'(x)- \frac{2x}{1+x^2}f(x)\right) = \frac{f'(x)}{1+x^2} - \frac{2xf(x)}{(1+x^2)^2}
$$
and
$$
(BTf)(x)=\frac{d}{dx}\left(\frac{f(x)}{1+x^2}\right)=  \frac{f'(x)}{1+x^2} - \frac{2xf(x)}{(1+x^2)^2}.
$$
{Thus, indeed, $TD(A) \subseteq D(B)$ and $TAf = BTf, \; \forall\, f \in D(A)$.
It is easily seen that $\sigma_p (A)=\emptyset$. As for $B$, one has, as it is well known, $\sigma(B)=\sigma_c(B)= i{\mb R}$. On the other hand, $0\in \sigma_r (A)$, since, if $h(x)=(1+x^2)^{-1}$, then $\ip{Af}{h}=0$, for every $f \in W^{1,2}({\mb R})$, so that the range $R(A)$ is not dense. Actually one has $\sigma_r(A)= \{0\}$, as one can easily check by computing $\sigma_p(A^*)$. Thus, by Corollary \ref{cor_3.24}, $\sigma(A)=\sigma(B)$, but the quasi-similarity does not preserve the relevant parts of the spectra.}
\enex

\section{The LHS generated by metric operators }
\label{sect_4}

Let $\M(\H)$ denote the family of all metric operators and $\M_b(\H)$ that of all bounded ones.
As said in Section \ref{sect_2},
 there is a natural order in $\M(\H) : G_1\preceq G_2$ iff $\H_{G_1}\subset \H_{G_2}$, where the embedding is continuous and has dense range.
  If $G_1,G_2$ are both bounded, a sufficient condition for $G_1\preceq G_2$ is that
 there exists $\gamma >0$  such that $G_2\leq \gamma G_1$.
Then  one has 
$$G_2^{-1} \preceq G_1^{-1} \Leftrightarrow G_1\preceq G_2 \; \mbox { if } G_1, G_2 \in \M_b(\H) 
$$
and 
$$
G^{-1} \preceq I \preceq G, \quad \forall \, G \in \M_b(\H).
$$

\berem
The family $\M(\H)$ is not necessarily directed upward with respect to $\preceq$. For instance, if $X, Y \in \M_b(\H)$, then $X,Y$ have the null operator 0 as a lower bound; 0
is not the greatest lower bound \cite[Section 2.8]{kadison}, but we cannot say that a positive lower bound exists. If a positive lower bound $Z$ exists, then $Z\in \M_b(\H)$ and by definition $X, Y \preceq Z$.
\enrem

As we will see now, the spaces $\{\H(X) : X \in \M(\H) \}$ constitute  a lattice of Hilbert spaces $V_{\I}$ in the sense of \cite[Definition 2.4.8]{pip_book}.

Let first $\O \subset \M(\H)$ be a family of metric operators 
and assume that
$$
\D:= \bigcap_{G\in \O} D(G^{1/2}) 
$$ 
is a dense subspace of $\H$. 
Of course, the condition is nontrivial only  if $\O$  contains unbounded elements, for instance, unbounded inverses of bounded operators.
We may always suppose that $I \in \O$.

Every operator $G\in \O$ is a self-adjoint, invertible operator. 
Then, on $\D$ we can define the graph topology ${\sf t}_{\O}$
by means of the norms
$$ 
\xi \in \D \mapsto \|G^{1/2}\xi\|, \quad G \in \O.
$$

Let $\D^\times$ be the conjugate dual of $\D[{\sf t}_{\O}]$, endowed with the strong dual topology ${\sf t}^\times_{\O}$.
Then the triplet
$$ 
\D[{\sf t}_{\O}] \hookrightarrow \H \hookrightarrow \D^\times[{\sf t}^\times_{\O}]
$$
is called the Rigged Hilbert Space  associated to $\O$.

 {As shown in \cite{interpolation}  and} in \cite[Section 5.5.2]{pip_book}, to the family $\O$ there corresponds a canonical lattice of Hilbert spaces (LHS).
 The lattice operations are defined by means  of the operators
\begin{align*} &X\wedge Y:= X \dotplus Y,\\ &X\vee Y:= (X^{-1} \dotplus Y^{-1})^{-1},
\end{align*}
where $\dotplus$ stands for the form sum and $X, Y \in \O$.
We recall that the form sum $T_{1 } \dotplus T_{2}$ of two positive operators is the positive (hence, self-adjoint) operator associated to the quadratic form $\t =\t_{1} + \t_{2}$,
where  $\t_{1}, \t_{2}$ are the quadratic forms of  $T_{1 }$ and $T_{2}$, respectively \cite[\S VI.2.5]{kato}.

 We notice that $X\wedge Y$  is  a metric operator,  but it need not belong to $ \O$.
First, it is self-adjoint and bounded from below by a positive quantity.
In addition, $(X\wedge Y)\xi=0$ implies $\xi=0, \, \forall\, \xi\in
Q(X \dotplus Y)= Q(X) \cap Q(Y)$, which is dense. Indeed, $\ip{(X+Y)\xi}{\xi} = \ip{X\xi}{\xi} +\ip{Y\xi}{\xi}=0$ implies
$\ip{X\xi}{\xi}=\ip{Y\xi}{\xi}=0$, since both $X$ and $Y$ are positive. This in turn implies $\xi=0$. Thus $X\wedge Y$ is  a metric operator, but it need not belong to $ \O$.
The same argument applies to the operator $X\vee Y$ .

In particular, if we take for  $\O$ the set $\M(\H)$ of all metric operators, we see that
it is stable under the lattice operations, i.e., it is a lattice by itself (but the corresponding domain $\D$ may fail to be dense).
This is not true in the general case envisaged in \cite[Section 5.5.2]{pip_book}.

For the corresponding Hilbert spaces, one has
\be\label{eq:lattice}
\begin{array}{rl} &\H(X\wedge Y):= \H(X) \cap \H(Y)\, ,
\\[1mm]
&\H(X\vee Y):= \H(X) + \H(Y)\, .
\end{array}
\en
 A second lattice, dual to the previous one, is obtained with the conjugate dual spaces $\H(X^{-1})$, as described in \cite[Section 5.5.2]{pip_book}.
The conjugate duals of the spaces \eqref{eq:lattice} are
\be\label{eq:lattice2}
\begin{array}{rl} &\H((X\wedge Y)^{-1}):= \H(X^{-1}) + \H(Y^{-1})\, ,
\\[1mm]
&\H((X\vee Y)^{-1}):= \H(X^{-1}) \cap \H(Y^{-1})\, .
\end{array}
\en

{Define the set $\R= \R(\O):= \{G^{\pm 1/2}, G \in \O\}$ and the corresponding domain $\D_\R:= \bigcap_{X\in \R} D(X)$.
Let now $\Sigma$ denote the minimal set of self-adjoint operators containing $\O$, stable under inversion and form sums, with the property 
that $\D_\R$ is dense
 in every $H_Z$, $Z \in \Sigma$ (i.e., $\Sigma$ is an admissible cone of self-adjoint operators \cite[Def. 5.5.4]{pip_book}). 
 Then, by \cite[Theorem 5.5.6]{pip_book}, $\O$ generates a lattice of \hs s $\I = \{\H(X), \, X \in \Sigma\}$ and a 
 \pip\ $V_\I$ with central Hilbert space $\H$ and total space $V=\sum_{G\in \Sigma}\H(G)$. The ``smallest'' space is $V^\#=\D_\R$.}
 The compatibility and the partial inner product read, respectively, as
\begin{align*}
 f \# g \;&\Longleftrightarrow\;  \exists \, G \in \Sigma  \; \mbox{ such that} \;f \in \H(G ),\, g \in \H(G^{-1}),
\\
\ip{f}{g}& = \ip {G^{1/2} f} { G^{-1/2} g}.
\end{align*}
{For instance, if $\O = \{I,G\}$, the set $\Sigma$ consists of the seven operators of Fig. \ref{fig:diagram3}, augmented by $I\wedge G \wedge G ^{-1}$ on the left and $I\vee G \vee G ^{-1}$ on the right. On the other hand, every power of $G$  is a metric operator.
Thus, if we take $\O = \{G ^{\alpha}, \alpha\in\ZN \; \textrm{or}\;\RN \}$, we obtain the scales $V_\J$ and $V_{\widetilde \J}$, 
 which are the \pip s generated by the construction above.
}

We denote by ${\rm Op}(V_\I)$ the space of operators in $V_\I$. As shown in \cite[Section 3.1.3]{pip_book}, an operator $A\in {\rm Op}(V_\I)$
can be described by the set ${\sf j}(A)$
 of pairs $(X,Y) \in \Sigma \times \Sigma $  such that $A$ maps $\H(Y)$ into $\H(X)$ continuously. We denote by
  $A\subn{XY} $   the $(X,Y)$-representative of $A$,
i.e., the restriction of $A$ to $\H(Y)$. Then $A$ is identified with the collection of its representatives:
$$ A\simeq \{A\subn{XY}: (X,Y)\in {\sf j}(A)\}.$$

Let us assume, in particular, that $(G,G) \in {\sf j}(A)$, for some {  $G \in \M(\H)$, bounded or not.} Then $A\subn{GG} $ is a bounded operator from $\H(G)$ into itself, i.e., there exists $c> 0$ such that
$$
\|G^{1/2}A\subn{GG} \xi\| \leq c\|G^{1/2}\xi\|, \quad \forall \,\xi \in \H(G).
$$
This means that
$$
\|G^{1/2}A\subn{GG}G^{-1/2} \eta\| \leq c\|\eta\|, \quad \forall\, \eta \in \H.
$$
Hence, ${\sf B}:=G^{1/2}A\subn{GG}G^{-1/2}$ is a bounded operator on $\H$.
Then the operator $A\subn{GG}\in {\mc B}(\H(G))$ is \emph{quasi-similar} to ${\sf B}\in \BH$, that is, $A\subn{GG}\dashv {\sf B} $.

More   generally, by an argument similar to that used above for the couple $(G,G)$, one can prove the following
\begin{prop} Let $A\in  {\rm Op}(V_\I)$. Then,
 $(X,Y)\in {\sf j}(A)$ if and only if $X^{1/2}AY^{-1/2}$ is a bounded operator in $\H$.
\end{prop}

\berem If the restriction ${\sf B}_{0}$ of ${\sf B}$ to $V^\#$ is continuous from $V^\#$ into $V$, then ${\sf B}_{0}$ determines a unique operator
 $B \in {\rm Op}(V_\I)$ \cite[Proposition 3.1.2]{pip_book} such that $B \xi = {\sf B}_{0}\xi$, for every $\xi \in V^\#$.
The previous statement then reads as follows: if $(G,G) \in {\sf j}(A)$, for some
 {  $G \in \M(\H)$,} then $(I,I)\in {\sf j}(B)$. \enrem

 {For $G \in \M_b(\H)$, {with unbounded inverse,} we have $\H \subset \H(G)$. Hence we can consider the restriction of
 $A\subn{GG}$ to $\H$, i.e., the operator ${\sf A}$  defined by}
\begin{align*}
D({\sf A})&= \{ \xi \in \H: \, A\subn{GG}\xi=A\xi \in \H\}\\
 {\sf A}\xi &= A\xi\, (=A\subn{GG}\xi), \quad \xi \in D({\sf A}) .
\end{align*}
In general $D({\sf A})$ need not be  dense in $\H$.

A sufficient condition for the density of $D({\sf A})$ can be given in terms of the {adjoint operator $A^\times$, which is defined by the relation
$$
\ip{A^\times x}{y} = \ip{x}{Ay},
$$
whenever the two partial inner products are well-defined (see \cite[Sec. 3.1.3]{pip_book} for a precise definition).}
 Indeed, by the assumption, $(G^{-1},G^{-1})\in  {\sf j}(A^\times)$.
 Then the space
$ D({\sf A}^\sharp):=\{\xi \in \H: \, A^\times\xi \in \H\}$ is dense in $\H$ since it contains $\H(G^{-1})$.
We define ${\sf A}^\sharp\xi =A^\times \xi$ for $\xi \in D({\sf A}^\sharp)$. Then we have

\beprop Let $(G,G) \in {\sf j}(A)$, with $G \in \M_b(\H)$. {Then the domain $D({\sf A})$ is dense in $\H$
 if and only if  the operator ${\sf A}^{\sharp\ast}$ is a restriction of $A$.}
  \enprop
 {\bf Proof. }  Let us assume first that ${\sf A}^{\sharp\ast}$ is a restriction of $A$. Then, since the domain  $D({\sf A})$ is maximal in $\H$,  we get that ${\sf A}^{\sharp\ast}={\sf A}$. Hence ${\sf A}$ is densely defined in $\H$.
{Conversely, for $f\in D({\sf A}^{\sharp\ast}) \subset \H \subset \H(G)$ and $ g\in \H(G^{-1})$, we have
\begin{align*}
\ip{{\sf A}^{\sharp\ast}f}{g} &= \ip{f}{{\sf A}^{\sharp}g} = \ip{f}{A^\times g} =  \ip{f}{A^\times\subn{G^{-1}G^{-1}} g}\\
 &= \ip{A\subn{GG}f}{g} = \ip{Af}{g},
\end{align*}
the last equality being valid on the dense domain  $D({\sf A})$.}
\qed

\section{Similarity for symmetric \pip\ operators}
\label{sect_5}

Let us come back to the \pip\ generated by the metric operators, described in Section \ref{sect_3}.
Given an operator $A\in {\rm Op}(V_\I)$, we define
 $$
{\sf s}(A)=\{X \in \Sigma :  (X,X)\in {\sf j}(A) \}.
$$
This set can be conveniently used to describe operators similar or quasi-similar to some representative of $A$.
 Notice that
$$
{\sf s}(A^\times)=\{X^{-1} : X \in{\sf s}(A)\} .
$$
From the definitions \eqref{eq:lattice}, it is clear that  the set ${\sf s}(A)$ is invariant under the lattice operations $\cap$ and $+$.
Coming back to the scale \eqref{eq:scale} or  its continuous extension $V_{\widetilde \J}:= \{\H_{\alpha}, \alpha \in \RN \}$,
associated to the fixed metric operator $G$,
we may identify  $\alpha \in \RN$ with $\H_{\alpha}= \H(G^{\alpha})$ and consider the subset
${\sf s}_{G}(A)=\{\alpha\in \RN : (\alpha,\alpha)\in  {\sf j}(A)  \} \subset \RN$.
Then ${\sf s}_{G}(A^\times) = \{-\alpha:  \alpha\in {\sf s}_{G}(A) \}$.
 \medskip

 Let $A\in  {\rm Op}(V_\I)$ and assume that $G\in {\sf s}(A)$ for $G \in \M(\H)$.
Then, as we said above,  the operator ${\sf B}:=G^{1/2}A\subn{GG}G^{-1/2}$ is bounded. Suppose now that $A\subn{GG}$ has a  restriction to $\H \subset \H(G)$, by which we mean, as before, that the subspace
$$
 D({\sf A})=\{ \xi \in \H:\, A\xi \in \H\}
$$
 is dense in $\H$ and the operator ${\sf A}:=A\upharpoonright D({\sf A})$ is closed.
Then we can define a second operator $\breve{\sf B}$ by
\begin{align*}
 D(\breve{\sf B}) & = \{\xi \in D(G^{-1/2}):\, G^{-1/2}\xi \in D({\sf A})\},  \\
 \breve{\sf B}\,\xi & = G^{1/2}{\sf A}G^{-1/2}\xi, \quad \xi \in D(\breve{\sf B}).
\end{align*}
It is clear that if $\eta \in D({\sf A})$, then $G^{1/2}\eta \in D(\breve{\sf B})$. The subspace $\M:= \{G^{1/2}\eta: \eta \in D({\sf A})\}$ is dense in $\H$ since,
if $\zeta \in \M^\perp$, one has
$$ \ip{G^{1/2} \zeta}{\eta}=\ip{\zeta}{G^{1/2}\eta}=0, \quad \forall\, \eta \in D({\sf A}).$$
The density of $D({\sf A})$, which has been assumed, implies that $G\zeta=0$ and then $\zeta=0$.
Therefore, if $D({\sf A})$ is dense, $D(\breve{\sf B})$ is also dense and $\breve{\sf B}$ is bounded since $\breve{\sf B}$ and ${\sf B}$  { coincide on $D(\breve{\sf B})$.}
This implies that $\overline{\breve{\sf B}}={\sf B}$. Moreover,  $G^{1/2}: D({\sf A}) \to D({\sf B})$ and
$$
 {\sf B} \,G^{1/2} \eta = G^{1/2}{\sf A}\, \eta, \quad \forall \,\eta \in D({\sf A}).
$$
 {Since $G^{1/2}$ is bounded, this means that} ${\sf A}\dashv {\sf B}$.

On the other hand,
$$
{\sf B} \,G^{1/2} \eta = G^{1/2}{A}\, \eta, \quad \forall \,\eta \in \H(G).
$$
So, if $G \in  {\sf s}(A)$, then $A$ is similar to a bounded operator in $\H$. But $G^{1/2}$ is a unitary operator from $\H(G)$ onto $\H$; hence $A$ and ${\sf B}$ are
 unitarily equivalent, while for the restriction  ${\sf A}$ of $A$ to $\H$ only quasi-similarity may hold.
 \medskip

{We come now to the case of symmetric operators, in the \pip\ sense, i.e., operators that satisfy  $A=A^\times$.} This is the class that can give rise to self-adjoint \emph{restrictions} to $\H$, thus candidates for observables in the case of the description of a quantum system.
One possible technique is the \pip\ version of the celebrated KLMN theorem, discussed at length in \cite[Section 3.3.5]{pip_book}.

Let  $A=A^\times$ be a symmetric operator. Then  $X\in {\sf s}(A)$ if, and only if $X^{-1}\in {\sf s}(A)$,
and this implies $I\in {\sf s}(A)$, by \cite[Cor. 3.3.24]{pip_book}.  In addition, since  ${\sf s}(A)$ is invariant under the lattice operations $\cap$ and $+$, it becomes a genuine (involutive) sublattice of $V_{\J}$.

 Let us consider an operator $G \in \M(\H)$, with $G\in {\sf s}(A)$.
Then $A$ fixes all three middle spaces in Fig. \ref{fig:diagram} and, therefore, all seven spaces of the lattice.
This applies, in particular, to
all three  spaces in the triplet \eqref{eq:triplet} if $G$ is bounded, or in the triplet   \eqref{eq:tri>1} or   \eqref{eq:tri<1} if $G$ is unbounded.  Moreover, by the interpolation property (iii) of \cite[Sec. 5.1.2]{pip_book},
$A$  leaves invariant every space $\H_{\alpha}, \alpha\in \ZN \mbox{ or } \RN$, in the scales $V_{\J}$ and $V_{\widetilde \J}$. In other words,
$A$ is a totally regular operator in these \pip s (see \cite[Def.3.3.12]{pip_book}), hence, ${\sf s}_{G}(A) = \ZN$ or $\RN$, respectively.
\medskip

Under these conditions,  $D({\sf A})$ is dense in $\H$. The corresponding operator ${\sf B}$ is
 unitarily equivalent to  (a restriction of) $A$,
  whereas ${\sf A}$ and ${\sf B}$ are quasi-similar. In conclusion,
\beprop
Every symmetric operator $A\in  {\rm Op}(V_\I)$ such that $G \in {\sf s}(A)$, with $G \in \M(\H)$, is quasi-similar to a bounded operator.
\enprop
\medskip
\noindent  {Given a closed densely defined operator $B$ in $\H$, one may ask whether there exist $A=A^\times\in  {\rm Op}(V_\I)$ such that $B$ is similar or quasi-similar to some representative of $A$. The question is open.}

However, the assumption that $G\in {\sf s}(A)$ is too strong for applications, since it implies that $A$ has a \emph{bounded} self-adjoint restriction to $\H$.
Assume instead that $(G^{-1},G)\in {\sf j}(A)$
for some $G \in \M_b(\H)$ with an unbounded inverse. Then $\H(G^{-1}) \subset \H(G)$ and we can apply the KLMN theorem in its \pip\ version, namely, Theorem 3.3.27 in \cite{pip_book}.
\beprop\label{prop-KLMN1}
Given a symmetric operator $A=A^\times$, assume there is a metric operator $G \in \M_b(\H)$ with an unbounded inverse, for which there exists  a $ \lambda \in \RN$ such that $A  - \lambda I$ has
an invertible representative $(A  - \lambda I)_{G G^{-1}}: \H(G^{-1}) \to \H(G).$
Then  $A\subn{G G^{-1}}$ has a unique restriction to a selfadjoint operator  {\sf A} in the
 Hilbert space $\H$. The number $\lambda$  does not belong to the spectrum of  {\sf A}.
  The (dense) domain of  {\sf A}  is  given by $D({\sf A})=\{ \xi \in \H:\, A\xi \in \H\}$.
The resolvent $({\sf A}  - \lambda)^{-1}$ is  compact (trace class, etc.) if and only if the natural embedding $ \H(G^{-1}) \to \H(G) $ is compact (trace class, etc.).
\enprop
We briefly recall the argument. Let $R_{G^{-1}G} = ((A  - \lambda I)\subn{G G^{-1}})^{-1} : \H(G) \to \H(G^{-1})$ be the bounded inverse of the assumed invertible representative.
{Define $R\subn{II}= E\subn{IG^{-1}}R_{G^{-1}G}E\subn{GI}$, which is a restriction of $R\subn{G^{-1}G}$
(here, as usual, $E\subn{YX}: \H(X) \to \H(Y) $ is the representative of the identity operator (embedding) when $\H(X) \subset\H(Y)$, and $\H(I):= \H$). Then, by the assumption,
$R\subn{II}$ is bounded and, by \cite[Lemma 3.3.26]{pip_book},
it has a self-adjoint inverse ${\sf A}-\lambda $, which is a restriction of $(A  - \lambda I)\subn{G G^{-1}}$. The rest is obvious.}
\medskip

Now this proposition can be generalized, by exploiting Theorem 3.3.28 in \cite{pip_book}, in the \pip\ language of Section \ref{sect_2}.
\beprop\label{prop-KLMN2}
Let $V_\J = \{\H_{n}, n\in \ZN\}$ be the Hilbert scale built on the powers of the  operator $G^{-1/2}$, where
$G \in \M_b(\H)$ with $G^{-1}$ unbounded.
Assume there is a $ \lambda \in \RN$ such that $A  - \lambda I$ has
an invertible representative $(A  - \lambda I)_{nm}: \H_m \to \H_n$, with $\H_{m} \subset \H_n$. Then the conclusions of Proposition \ref{prop-KLMN1} hold true.
\enprop
According to the proof of \cite[Theorem 3.3.28]{pip_book}, the assumption implies that the operator $R_{mn}= (A_{mn}  - \lambda I_{mn})^{-1}: \H_n \to \H_m$
has a self-adjoint representative $R\subn{II}$ in $\H$, which is injective and has dense range. Therefore, its inverse
${\sf A}- \lambda I= R\subn{II}^{-1}$, thus also {\sf A} itself, is defined on a dense domain and is self-adjoint.

The same result holds true in the case of an unbounded operator $G$, using the scale  \eqref{eq:scale} built on the powers of  $G^{1/2}$ or  $R_G^{1/2}$. Thus globally, we may  state
\beprop\label{prop-KLMN3}
Let $V_\J= \{\H_{n}, n\in \ZN\}$ be the Hilbert scale built on the powers of the  operator $G^{\pm1/2}$ or
 $R_G^{1/2}$, depending on the (un)boundedness of $G \in \M(\H)$. Assume there is a $ \lambda \in \RN$ such that
 $A  - \lambda I$ has
an invertible representative $(A  - \lambda I)_{nm}: \H_m \to \H_n$, with $\H_{m} \subset \H_n$. Then the conclusions of Proposition \ref{prop-KLMN1} hold true.
\enprop

{At this stage, we do have a self-adjoint restriction ${\sf A}$  of $A$ in $\H$, but we don't know if
 there any quasi-similarity relation between  $A\subn{G G^{-1}}$ or ${\sf A}$ and another operator. }

So far we have considered only the case of one metric operator $G$ in relation to $A$. Assume now we take two different metric operators $G_{1}, G_{2}\in \M(\H)$.
What can be said concerning $A$ if it maps $\H(G_{1})$ into $ \H(G_{2})$?
 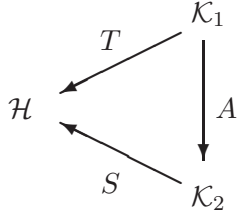
\begin{figure}[t]
\centering \setlength{\unitlength}{0.5cm}
\begin{picture}(5,4 )

\put(4,3){\begin{picture}(5,4 ) \thicklines

 \put(-1.6,2){\vector(-2,-1){3.2}}
 \put(-1.6,-2){\vector(-2,1){3.2}}
 \put(-1,1.8){\vector(0,-1){3.2}}

 \put(-6,0){\makebox(0,0){ $\H $}}
\put(-1,2.5){\makebox(0,0){ $\K_{1} $}}
\put(-1,-2.4){\makebox(0,0){ $\K_{2}$}}

 \put(-0.4,0){\makebox(0,0){$A$}}
\put(-3.6,1.8){\makebox(0,0){ $T$}}
\put(-3.6,-2){\makebox(0,0){ $S$}}
\end{picture}}
\end{picture}
\caption{\label{fig:semisim}The semi-similarity scheme.}
\end{figure}

One possibility is to introduce a notion slightly more general than quasi-similarity, called \emph{semi-similarity}.
\bedefi
Let $\H, \,\K_{1}$ and $ \K_{2}$ be three Hilbert spaces,  $A$ a closed, densely defined operator from $\K_{1}$ to  $\K_{2}$, $B$ a closed, densely defined operator on $\H$. Then $A$ is said to be
\emph{semi-similar} to $B$, which we denote by $A\dashvv B$, if there exist two bounded operators  $T:\K_{1}\to \H$ and $S:\K_{2}\to \H$ such that (see Fig. \ref{fig:semisim}):
\begin{itemize}
\item[(i)] $T:D(A)\to D(B)$;
\item[(ii)] $BT\xi = SA\xi, \; \forall\, \xi \in D(A)$.
\end{itemize}
The pair $(T,S)$ is called an \emph{intertwining couple}.
\findefi
Of course, if  $\K_{1}=\K_{2}$ and $S=T$, we recover the notion of quasi-similarity and $A\dashv B$.

Now we come back to the case envisaged above: $A:\H(G_{1}) \to \H(G_{2})$ continuously, for the two metric operators $G_{1}, G_{2}\in \M_b(\H)$, but $A$ is not supposed to be symmetric.
Under this assumption, we essentially recover the previous situation.  Since $A\subn{G_{2}G_{1}}$ is a bounded operator from $\H(G_{1})$ into $\H(G_{2})$,  there exists $c>0$ such that
$$
\|G_{2}^{1/2}A\subn{G_{2}G_{1}} \xi\| \leq   c\|G_{1}^{1/2}\xi\|, \quad \forall\, \xi \in \H(G_{1)}.
$$
This means that
$$
\|G_{2}^{1/2}A\subn{G_{2}G_{1}}G_{1}^{-1/2} \eta\| \leq  c \|\eta\|, \quad \forall\, \eta \in \H.
$$
Hence, ${\sf B}:=G_{2}^{1/2}A\subn{G_{2}G_{1}}G_{1}^{-1/2} $ is bounded in $\H$.
Then the operator $A\subn{G_{2}G_{1}}$ is \emph{semi-similar} to ${\sf B}\in \BH$, that is, $A\subn{G_{2}G_{1}}\dashvv {\sf B} $,
with respect to the intertwining couple  $T=G_{1}^{1/2}, S=G_{2}^{1/2} $.

Next we take   $A=A^\times$   symmetric. Then  $A:\H(G_{1}) \to \H(G_{2})$ continuously implies  $A:\H(G_{2}^{-1}) \to \H(G_{1}^{-1})$ continuously as well.
Assume that $G_{1} \preceq G_{2}$, that is,  $\H(G_{1}) \subset \H(G_{2}) $. This yields the following situation:
$$
\H(G_{2}^{-1}) \;\subset\;  \H(G_{1}^{-1}) \;\subset\;  \H \;\subset\;  \H(G_{1})\;\subset\;  \H(G_{2}).
$$
Therefore, since $\H(G_{1}^{-1}) \hookrightarrow \H(G_{2})$, the operator $A$ maps $\H(G_{2}^{-1})$ continuously into $\H(G_{2})$, that is,
we are back to the situation of Proposition \ref{prop-KLMN1} and we can state:

\beprop\label{prop-KLMN4}
Given a symmetric operator $A=A^\times\in  {\rm Op}(V_\I)$, assume there exists two metric operators $G_{1},G_{2}\in \M_b(\H)$ such that
$G_{1}\preceq G_{2}$ and  $(G_{1},G_{2})\in {\sf j}(A)$.
Assume   there exists  a $ \lambda \in \RN$ such that $A  - \lambda I$ has  an invertible representative
$(A  - \lambda I)_{G_{2} G_{2}^{-1}}: \H(G_{2}^{-1}) \to \H(G_{2}).$
Then there exists a unique restriction of $A\subn{G_{2} G_{2}^{-1}}$ to a self-adjoint operator  {\sf A} in the   Hilbert space $\H$ and the other  conclusions of Proposition \ref{prop-KLMN1}
hold true.
\enprop
The analysis may be extended to the three other cases, assuming again that $A:\H(G_{1}) \to \H(G_{2})$:
\begin{enumerate}
\item[(i)] $G_1$ bounded, $G_2$ unbounded: then
$$
\H(G_{2})  \subset \H  \subset \H(G_{1}) \quad \mbox{and}
\quad \H(G_{1}^{-1})  \subset \H   \subset \H(G_{2}^{-1}),  
$$
so that, in both cases,  $A$ maps the large  space into the small  one  and, therefore, the KLMN theorem \emph{does not apply}.
\item[(ii)]   $G_1$ unbounded, $G_2$ bounded: then
$$
\H(G_{1}) \; \subset\; \H  \; \subset\; \H(G_{2})
$$
and $A$ maps the small space into the large one. Again the KLMN theorem applies without restriction.
\end{enumerate}
In conclusion, if $A=A^\times$ is  symmetric and $(G_{1},G_{2}) \in {\sf j}(A)$, the KLMN theorem applies and yields a self-adjoint restriction in $\H$ in three cases: 
\begin{enumerate}
\item[(i)] If $G_{1}$ and $G_{2}$ are both bounded or  both unbounded, the theorem applies in the appropriate infinite scale, \emph{provided}
 $\H(G_{1}) \subset \H(G_{2})$.
 
 \item[(ii)]   If $G_1$ is unbounded and $G_2$ is bounded, the inclusion is automatic, 
thus the theorem applies without restriction.
\end{enumerate}
 On the other hand, if  $G_1$ is bounded and  $G_2$ is unbounded, the inclusion $\H(G_{1}) \subset \H(G_{2})$ cannot take place, hence the theorem does not apply.

\section{The case of pseudo-hermitian operators}
\label{sect_6}

Metric operators appear routinely in the so-called pseudo-hermitian quantum mechanics \cite{bender}, but in general only bounded ones are considered. In some recent work \cite{bag-zno,mosta2}, however, unbounded metric operators have been discussed. The question is, how do these operators fit in the present formalism?

Following the argument of \cite{mosta2}, the starting point   is a reference \hs\ $\H$ and a $G$-pseudo-hermitian operator $H$ on $\H$, which means  there exists an unbounded metric operator $G$ satisfying the relation
\be\label{eq:psH}
H^\ast G = G H.
\en
In the relation \eqref{eq:psH}, the two operators are assumed to have the same dense domain, $D(H^\ast G)=D(GH)$.
Such an operator $H$, which is the putative non-self-adjoint (but $\P\T$-symmetric) Hamiltonian of a quantum system, is also called \emph{quasi-hermitian} \cite{dieudonne}.

Next one assumes that the operator $H$ possesses a (large) set of vectors, $\D:=\D_G^\omega(H)$, which are  analytic in the norm  $\norm{G}{\cdot}$ and are contained in $D(G)$  \cite{barut-racz,nelson}.
This means that every vector  $\phi\in\D_G^\omega(H)$ satisfies the relation
$$
\sum_{n=0}^{\infty}\frac{ \norm{G}{H^n \phi}}{n!} \, t^n < \infty, \mbox{ for some }  t\in \RN.
$$
Then one endows $\D$ with the norm $\norm{G}{\cdot}$ and takes the completion $\H_G$, which is a closed subspace of $\H(G)$, as defined in Section \ref{sect_2}. An immediate calculation then yields
$$
\ip{\phi}{H\psi}_G = \ip{H\phi}{\psi}_G, \; \forall\, \phi,\psi\in \D,
$$
that is, $H$ is a densely defined symmetric operator in $\H_G$. Since it has a dense set of analytic vectors, it is essentially self-adjoint, by Nelson's theorem \cite{barut-racz,nelson}, hence its closure $\ov H$ is a self-adjoint operator in $\H_G$.  The pair $(\H_G, \ov H)$ is then interpreted as the physical quantum system.

{Next, by definition, $W_\D:= G^{1/2}\up \D$ is isometric from  $\D$ into $\H$, hence it extends to an isometry 
$W =\ov{W_\D}: \H_G \to \H$.  The range of the latter  is a closed subspace of $\H$,  denoted  $\H_{\rm phys}$, and the operator $  W$ is unitary  from $\H_G$ onto $\H_{\rm phys}$. Therefore, the operator $h=  W\, \ov H\,W^{-1}$ is self-adjoint in $\H_{\rm phys}$. This operator $h$ is interpreted as the genuine Hamiltonian of the system, acting in the physical \hs\  $\H_{\rm phys}$.}

{Things simplify if $\D$ is dense in $\H$: then $ W(\D)$ is also dense, $\H_G= \H(G)$, 
 $\H_{\rm phys}= \H$ and $W = G^{1/2}$ is unitary from $\H(G)$ onto $ \H$.
Also, if $G$ is bounded, it is sufficient to  assume that the vectors in $\D$ are analytic with respect to the original norm of $\H$.}

Now, every eigenvector of an operator is automatically analytic, hence this construction generalizes that of \cite{mosta2}.
This applies, for instance, to the example given there, namely, the $\P\T$-symmetric operator
$H=\frac12 (p-i\alpha)^2 + \frac12 \omega^2 x^2$  in $\H= L^2(\RN)$, for any $\alpha\in \RN$, which has an orthonormal basis of eigenvectors.

\section{Conclusion}

We have seen that the consideration of unbounded metric operators leads naturally to the formalism of \pip s.  On the other hand, we have introduced  several generalizations of similarity between operators and we have obtained some results on the preservation of spectral properties under quasi-similarity, but only with a bounded metric operator with unbounded inverse.
Then it turns out that exploiting the connection between metric operators and \pip s  may improve the quasi-similarity of operators.

Of course, these results are only a first step, many open problems subsist. In view of the applications, notably
in pseudo-hermitian QM, the most crucial ones concern the behavior of spectral properties under some generalized similarity with an unbounded metric operator. Research in this direction is in progress.

\end{document}